\documentclass[preprint,review,12pt]{elsarticle}

\usepackage{graphics}
\usepackage{color}

\usepackage{amssymb}

\usepackage{lineno}

\journal{Icarus}

\begin{document}

\begin{frontmatter}

\title{On the possible noble gas deficiency of Pluto's atmosphere}

\author{Olivier~Mousis}
\ead{olivier.mousis@obs-besancon.fr}
\address{Universit{\'e} de Franche-Comt{\'e}, Institut UTINAM, CNRS/INSU, UMR 6213, Besan\c con Cedex, France}
\address{Universit\'e de Toulouse; UPS-OMP; CNRS-INSU; IRAP; 14 Avenue Edouard Belin, 31400 Toulouse, France}
\author{Jonathan I. Lunine}
\address{Center for Radiophysics and Space Research, Space Sciences Building Cornell University,  Ithaca, NY 14853, USA}
\author{Kathleen E. Mandt}
\address{Space Science and Engineering Division, Southwest Research Institute, San Antonio, TX 78228, USA}
\author{Eric~Schindhelm}
\address{Southwest Research Institute, 1050 Walnut Street, Boulder, CO 8030223, USA}
\author{Harold~A.~Weaver}
\address{Space Department, Johns Hopkins University Applied Physics Laboratory, 11100 Johns Hopkins Road, Laurel, MD 20723-6099, USA}
\author{S. Alan~Stern}
\address{Southwest Research Institute, 1050 Walnut Street, Boulder, CO 8030223, USA}
\author{J. Hunter Waite}
\address{Space Science and Engineering Division, Southwest Research Institute, San Antonio, TX 78228, USA}
\author{Randy Gladstone}
\address{Space Science and Engineering Division, Southwest Research Institute, San Antonio, TX 78228, USA}
\author{Audrey Moudens}
\address{LERMA, Universit{\'e} de Cergy-Pontoise, Observatoire de Paris, ENS, UPMC, UMR 8112 du CNRS, 5 mail Gay Lussac, 95000 Cergy Pontoise Cedex, France}

\begin{abstract}
{We use a statistical-thermodynamic model to investigate the formation and composition of noble-gas-rich clathrates on Pluto's surface.} By considering an atmospheric composition close to that of today's Pluto and a broad range of surface pressures, we find that Ar, Kr and Xe can be efficiently trapped in clathrates if they formed at the surface, in a way similar to what has been proposed {for} Titan. The formation on Pluto of clathrates rich in noble gases could then induce a strong decrease in their atmospheric abundances relative to {their} initial values. A clathrate thickness {of  order of a few centimeters} globally averaged on the planet is enough to trap all Ar, Kr and Xe if these noble gases were in protosolar proportions in {Pluto's early} atmosphere. Because atmospheric escape over an extended period of time (millions of years) should lead to a noble gas abundance that either remains constant or increases with time, we find that a potential depletion of Ar, Kr and Xe in the atmosphere would best be explained by their trapping in clathrates. A key observational test is the measurement of Ar since the Alice UV spectrometer aboard the New Horizons spacecraft will be sensitive enough to detect its abundance $\sim$10 times smaller than in the case considered here.
\end{abstract}


\end{frontmatter}

\linenumbers

\section{Introduction}
\label{intro}

Gas hydrates, or clathrates, may exist throughout the solar system. Comparison of predicted stability fields of clathrates with conditions in various planetary environments suggest that these structures could be present in the Martian permafrost (Musselwhite \& Lunine 1995; Thomas et al. 2009; Swindle et al. 2009; Herri \& Chassefi{\`e}re 2012; Mousis et al. 2013), on the surface and in the interior of Titan (Tobie et al. 2006; Mousis \& Schmitt 2008), and in other icy satellites (Prieto-Ballesteros et al. 2005; Hand et al. 2006). It has also been suggested that the activity of many comets could result from the destabilization of these ices (Marboeuf et al. 2010, 2011, 2012a). On Earth, the destabilization of significant masses of CO$_2$ and CH$_4$ potentially stored in clathrates buried in seabeds and permafrost is regarded as a possible aggravating factor in future global warming (clathrate gun hypothesis -- Kennett et al. 2003). Broadly speaking, clathrates are thought to have taken part in the assemblage of the building blocks of many bodies of the solar system and may be in other planetary systems (Lunine \& Stevenson 1985; Mousis et al. 2002, 2006, 2009, 2010, 2011, 2012a; Mousis \& Gautier 2004; Alibert et al. 2005; Marboeuf et al. 2008; Madhusudhan et al. 2011; Johnson et al. 2012).

Clathrates have also been proposed to be at the origin of the noble gas deficiency measured in situ by the Huygens probe in the atmosphere of Titan (Osegovic \& Max 2005; Thomas et al. 2007, 2008; Mousis et al. 2011). In the case of Mars, important quantities of argon, krypton and xenon are believed to be trapped in clathrates located in the near subsurface and their storage in these structures could explain the measured two order of magnitude drop between the noble gas atmospheric abundances in Earth and Mars (Mousis et al. 2012b). Here we investigate the possibility of formation of clathrates rich in noble gases on Pluto's surface. To do so, we use the same statistical-thermodynamic model applied to the case of Titan to determine the composition of clathrates that might form on Pluto {and to investigate the possible consequences of their presence on the atmospheric composition.}

\section{The statistical--thermodynamic model}
\label{model}

To calculate the relative abundances of guest species incorporated in a multiple guest clathrate (hereafter MG clathrate) at given temperature and pressure, we use a model applying classical statistical mechanics that relates the macroscopic thermodynamic properties of clathrates to the molecular structure and interaction energies (van der Waals \& Platteuw 1959; Lunine \& Stevenson 1985). It is based on the original ideas of van der Waals and Platteeuw for clathrate formation, which assume that trapping of guest molecules into cages corresponds to the three-dimensional generalization of ideal localized adsorption.
	
In this formalism, the fractional occupancy of a guest molecule $K$ for a given type $t$ ($t$~=~small or large) of cage (see Sloan 1998; Sloan \& Kohn 2008) can be written as

\begin{equation}
y_{K,t}=\frac{C_{K,t}P_K}{1+\sum_{J}C_{J,t}P_J} ,
\label{eq1}
\end{equation}

\noindent where the sum in the denominator includes all the species which are present in the initial gas phase. $C_{K,t}$ is the Langmuir constant of species $K$ in the cage of type $t$, and $P_K$ is the partial pressure of species $K$. This partial pressure is given by $P_K=x_K\times P$ (we assume that the sample behaves as an ideal gas), with $x_K$ the mole fraction of species $K$ in the initial gas, and $P$ the total atmospheric gas pressure, which is dominated by N$_2$. The Langmuir constant depends on the strength of the interaction between each guest species and each type of cage, and can be determined by integrating the molecular potential within the cavity as

\begin{equation}
C_{K,t}=\frac{4\pi}{k_B T}\int_{0}^{R_c}\exp\Big(-\frac{w_{K,t}(r)}{k_B T}\Big)r^2dr ,
\label{eq2}
\end{equation}

\noindent where $R_c$ represents the radius of the cavity assumed to be spherical, $k_B$ the Boltzmann constant, and $w_{K,t}(r)$ is the spherically averaged Kihara potential representing the interactions between the guest molecules $K$ and the H$_2$O molecules forming the surrounding cage $t$. This potential $w(r)$ can be written for a spherical guest molecule, as (McKoy \& Sinano\u{g}lu 1963)

\begin{eqnarray}
w(r) 	= 2z\epsilon\Big[\frac{\sigma^{12}}{R_c^{11}r}\Big(\delta^{10}(r)+\frac{a}{R_c}\delta^{11}(r)\Big) \\ \nonumber
- \frac{\sigma^6}{R_c^5r}\Big(\delta^4(r)+\frac{a}{R_c}\delta^5(r)\Big)\Big],
\label{eq3}
\end{eqnarray}

\noindent with

\begin{equation}
\delta^N(r)=\frac{1}{N}\Big[\Big(1-\frac{r}{R_c}-\frac{a}{R_c}\Big)^{-N}-\Big(1+\frac{r}{R_c}-\frac{a}{R_c}\Big)^{-N}\Big].
\label{eq4}
\end{equation}

\noindent In Eq. 3, $z$ is the coordination number of the cell. This parameter depends on the structure of the clathrate (I or II;  see Sloan \& Koh 2008) and on the type of the cage (small or large). The Kihara parameters $a$, $\sigma$ and $\epsilon$ for the molecule-water interactions, given in Table \ref{Kihara}, have been taken from the recent compilation of Sloan \& Koh (2008) when available and from Parrish \& Prausnitz (1972) for the remaining species.

Finally, the mole fraction $f_K$ of a guest molecule $K$ in a clathrate can be calculated with respect to the whole set of species considered in the system as

\begin{equation}
f_K=\frac{b_s y_{K,s}+b_\ell y_{K,\ell}}{b_s \sum_J{y_{J,s}}+b_\ell \sum_J{y_{J,\ell}}},
\label{eq5}
\end{equation}

\noindent where $b_s$ and $b_l$ are the number of small and large cages per unit cell respectively, for the clathrate structure under consideration, and with ${\sum_{K}} f_{K}~=~1$. Values of $R_c$, $z$, $b_s$ and $b_l$ are taken from Parrish \& Prausnitz (1972).

In the present approach, the dissociation pressure of the MG clathrate and the mole fractions of the trapped volatiles are independently calculated. All mole fraction calculations are performed at the dissociation pressure $P =  P^{diss}_{mix}$ of the clathrate, i.e. temperature and pressure conditions at which this ice forms. This dissociation pressure can be deduced from the dissociation pressure $P_K^{diss}$ of a pure clathrate of species $K$, as (Hand et al. 2006; Thomas et al. 2007):

\begin{equation}
P^{\rm diss}_{\rm mix}=\Big(\sum_{K}\frac{x_K}{P^{\rm diss}_K}\Big)^{-1},
\label{eq6}
\end{equation}

\noindent where $x_K$ is the atmospheric mole fraction of species $K$ and $P^{\rm diss}_K$ its dissociation pressure. $P^{\rm diss}_K$ derives from laboratory measurements and follows an Arrhenius law (Miller 1961) as

\begin{equation}
\log(P^{\rm diss})=A+\frac{B}{T} ,
\label{eq7}
\end{equation}

\noindent where $P^{\rm diss}$ is expressed in Pa and $T$ is the temperature in K. {The constants $A$ and $B$ for N$_2$ and Ar have been fitted to experimental data (Lunine \& Stevenson 1985; Sloan 1998) and those for Xe and Kr have been taken from Fray et al. (2010) and Marboeuf et al. (2012b), respectively (see Table \ref{tab:coeff_pression}).} 

\section{Clathrate production on Pluto}
\label{results}

Depending on its composition, the dissociation pressure of MG clathrate varies between $\sim$5.1~$\times$~10$^{-10}$ and 2.1~$\times$~10$^{-5}$ Pa at Pluto's average surface temperature ($\sim$50 K; Lellouch et al. 2000), a value well below the atmospheric surface pressure which lies in the 0.65--2.4 Pa range (Elliot et al. 2003; Sicardy et al. 2003). This implies that MG clathrate remains stable at the surface irrespective of Pluto's seasonal variations and that its dissociation and reformation cannot occur under the planet's {current atmospheric conditions}. The only conditions on Pluto's surface allowing MG clathrate formation from an initial inventory present in the atmosphere require that the surface be hotter than {the present-day temperature}, most likely at early epochs after the planet's formation (see e.g. Fig. 1 of McKinnon 2002) or during the collisions that engendered the Pluto--Charon binary system (McKinnon 1988, 1989). An alternative possibility would be the recent or ancient release of hot ice from the interior of Pluto as the result of cryovolcanic events (Cook et al. 2007). It is during clathrate formation that substantial amounts of volatiles might have been sequestrated from the atmosphere into the surface.

\begin{figure}
\begin{center}
\includegraphics[angle=0,width=13cm]{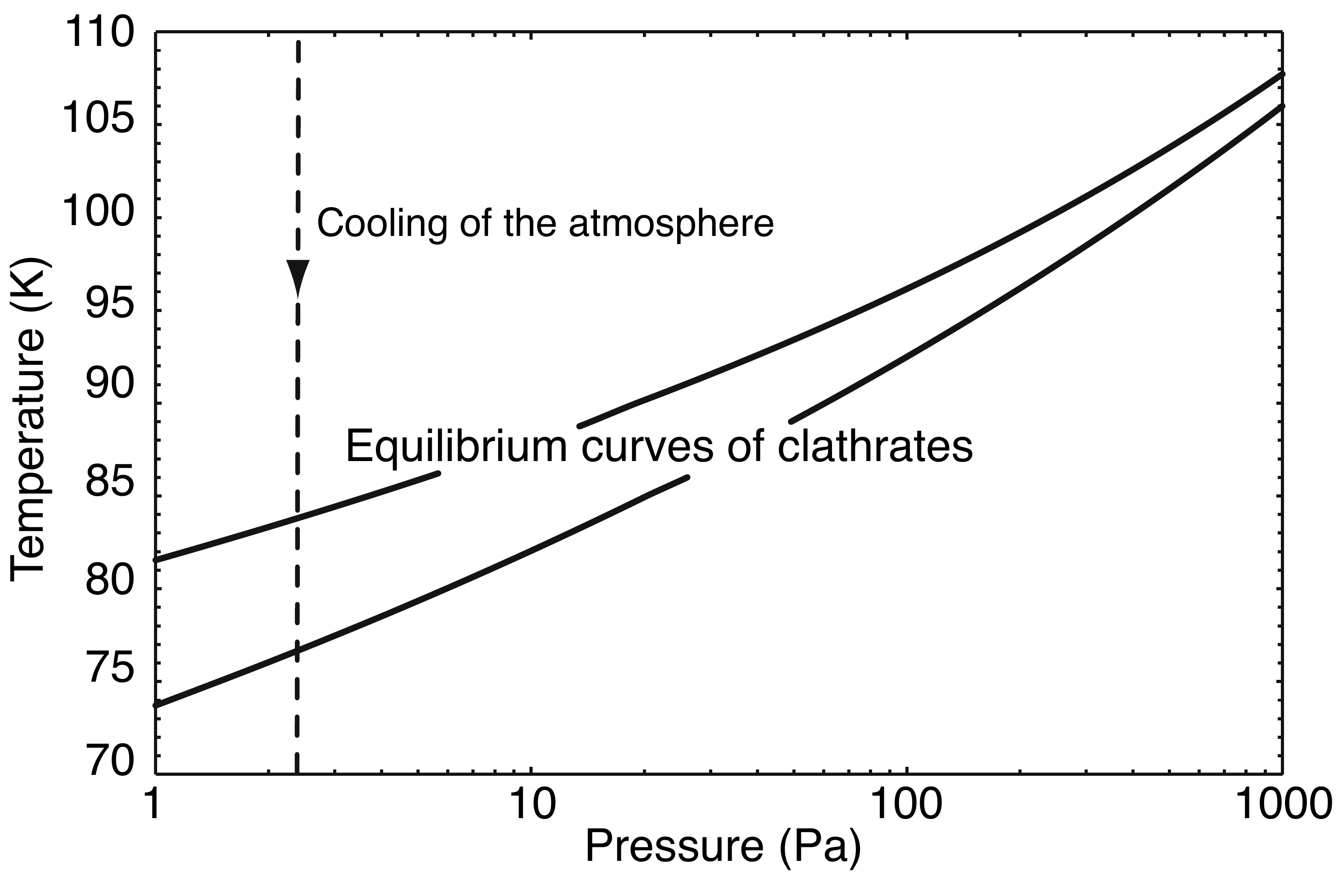}
\caption{From top to bottom: equilibrium curves of Xe-, Kr-, and Ar-dominated clathrates (the latter two appear superimposed). The arrow pointing down represents the path followed by a cooling atmosphere with a surface pressure of 2.4 Pa. When the cooling curve intercepts an equilibrium curve, then the corresponding clathrate forms.} 
\label{stab}
\end{center}
\end{figure}

Table \ref{atm} gives the composition of Pluto's proto-atmosphere used in our calculations. We made the conservative assumption that all noble gases were initially present in the proto-atmosphere of Pluto, with Ar/N, Kr/N and Xe/N ratios assigned to be protosolar (Asplund et al. 2009). Because Pluto's proto-atmosphere is expected to be strongly dominated by N$_2$ and that N$_2$ clathrate is of structure II (Lunine \& Stevenson 1985; Sloan \& Koh 2008), we show here calculations of the MG clathrate composition only for this structure. It is also important to note that our composition calculations are only valid along the dissociation curve of the clathrate of interest {(see Fig. \ref{stab})}.

\begin{figure}
\begin{center}
\includegraphics[angle=-90,width=10cm]{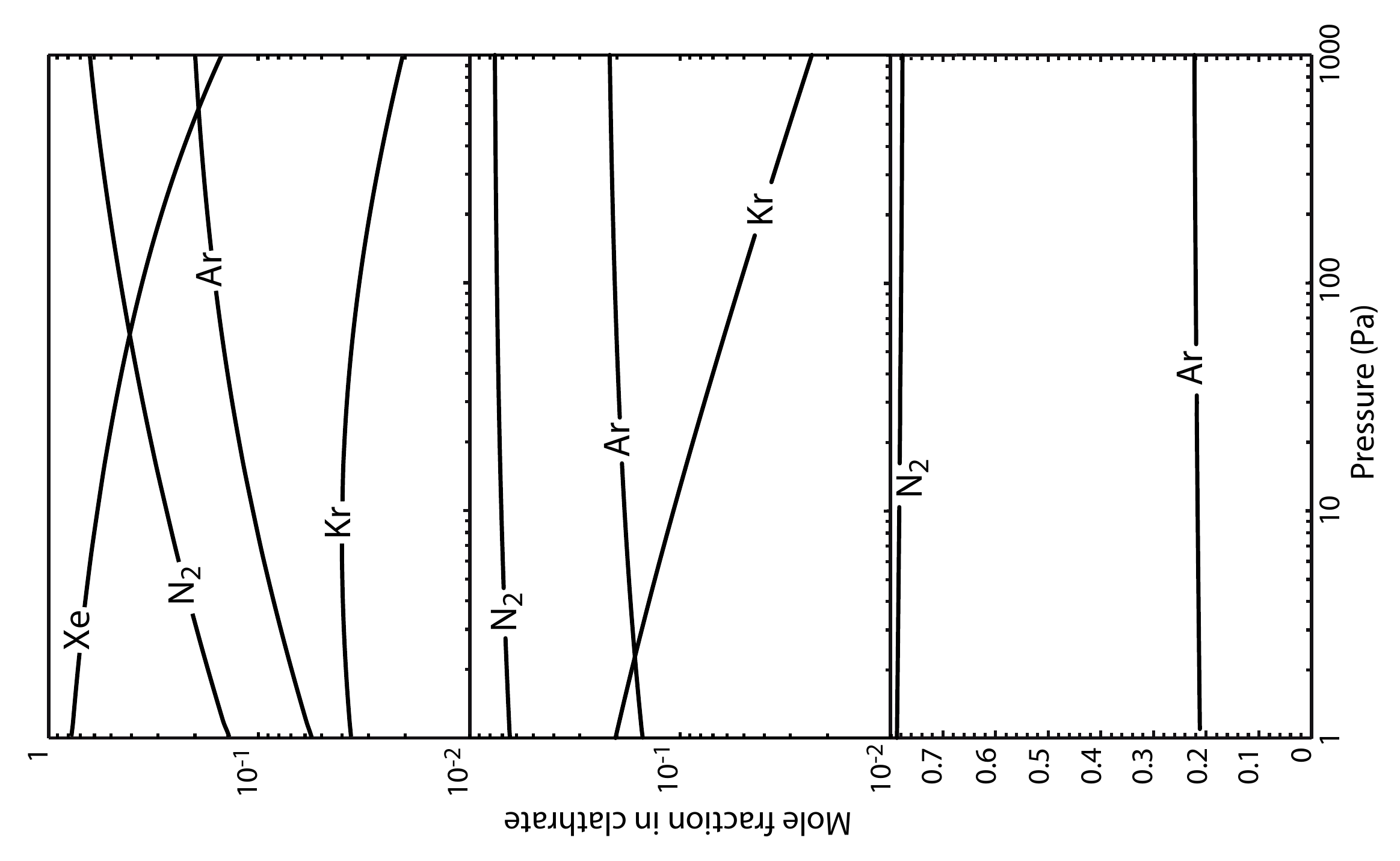}
\caption{{From top to bottom: mole fractions of volatiles encaged in clathrates that successively form} and calculated as a function of the surface pressure of N$_2$. Noble gas abundances are assumed to be protosolar relative to N$_2$ (Asplund et al. 2009). The clathrate composition is investigated from a gas phase composed of Ar, Kr, Xe and N$_2$ (top panel), of Ar, Kr and N$_2$ (middle panel), and of Ar and N$_2$ (bottom panel).} 
\label{mole}
\end{center}
\end{figure}

{From our calculations, we find that three kinds of clathrates with distinct compositions might form on Pluto's surface, each of them containing noble gases in different proportions.} Figure \ref{mole} shows the composition of these clathrates computed for an atmospheric pressure ranging between 1 and 10$^3$ Pa. Note that these {surface} pressures correspond to a MG clathrate equilibrium temperature in the $\sim$80--100 K range. {If a greater surface pressure is considered, then the MG clathrate will also form at a higher equilibrium temperature}. A first layer forms from the gas phase composition depicted in Table \ref{atm}. Irrespective of the pressure considered, the mole fraction of Xe trapped in this clathrate is between {$\sim$0.15 and 0.76, i.e. a range of values that is $\sim$31,000--159,000} times larger than its atmospheric mole fraction. The mole fraction of Kr is also substantially enhanced by a factor of {$\sim$400--750} in clathrate compared to its atmospheric value. {In contrast, the mole fraction of Ar trapped in clathrate evolves from a slight impoverishment ($\sim$0.5) to a moderate enrichment ($\sim$3) with increasing pressure, compared to its atmospheric value.} The second clathrate layer forms once Xe is fully trapped in the first layer. In this case, only N$_2$, Kr and Ar remain in the gas phase (the mole fractions of these species trapped in the first layer remain negligible). The mole fraction of Kr trapped in this clathrate is enhanced by a factor of {400--4,000 times} compared to its atmospheric mole fraction in the 1--10$^3$ Pa pressure range. The Ar mole fraction in this layer is also found to be moderately enriched by a factor of {$\sim$2--3}, compared to its atmospheric mole fraction. In its turn, a third clathrate layer forms when Kr is fully trapped in the second layer and in this case only N$_2$ and Ar remain in the coexisting gas phase. The fraction of Ar trapped in this clathrate remains constant irrespective of the  surface pressure considered and is found to be {$\sim$3 times} larger than its atmospheric value.

If Ar, Kr and Xe were initially in protosolar abundances in the atmosphere of Pluto, the amount of clathrates needed for their sequestration is relatively low. For example, if the three clathrate layers formed at their equilibrium temperatures\footnote{The equilibrium temperatures of the first, second and third clathrate layers are $\sim$77, 76 and 76 K, respectively.} for a surface pressure of 2.4 Pa, their total equivalent thickness is of order {$\sim$4 mm globally averaged on the planet}, assuming a full clathration efficiency and the presence of a structure II clathrate. {Interestingly, calculations conducted in the case of formation of structure I clathrate lead to similar conclusions. The noble gas trapping efficiencies are still very high but require an overall thickness of the three clathrate layers of $\sim$2 cm. In both clathrate structures, the equivalent layer of the Ar-dominated clathrate is more than 3 orders of magnitude thicker than those of Xe- and Kr-dominated clathrates. One must also note that the noble-gas-rich clathrates formed on Pluto could consist in a mixture of structures I and II clathrates. Indeed, in our computations of the composition of structure II clathrates, we find that the first layer formed is dominated by Xe, which is itself predicted to form a structure I clathrate (Sloan \& Koh 2008).}

\section{Competition with atmospheric escape}
\label{compet}

An alternative method for losing these noble gases from Pluto's atmosphere could be atmospheric escape. The atmosphere of Pluto is expected to escape efficiently due to Pluto's low gravity (Strobel 2008), though debate exists as to whether the escape rate is greater than subsonic (Tucker et al. 2012). If the escape of Pluto's atmosphere is hydrodynamic (Trafton et al. 1997; Strobel 2008), the outflow of N$_2$ provides enough energy to drag Ar, Kr and Xe from the atmosphere. The escape rate of these gases depends directly on the escape rate of N$_2$ {(Hunten et al. 1987)}:

\begin{equation}
F_i = \frac{X_i}{X_{N_2}} F_{N_2} \left (\frac{m_c - m_i}{m_c - m_{N_2}} \right ),
\label{eq8}
\end{equation}

\noindent where $F$ is the escape flux, $X$ is the abundance of the noble gas (subscript $i$) and N$_2$, $m$ is the mass of each constituent and $m_c$ is the critical mass which is also a function of the N$_2$ escape rate {(Hunten et al. 1987)}:

\begin{equation}
m_c = m_{N_2} + \frac{k~T~F_{N_2}}{b~g~X_{N_2}},
\label{eq9}
\end{equation}

\noindent where $k$ is the Boltzmann's constant, $T$ is the temperature, $b$ is the binary diffusion coefficient given as $b = AT^s$ with $A$ and $s$ constants determined through modeling or laboratory measurements, and $g$ is the acceleration due to gravity.

It is clear from Eq. \ref{eq9} that the critical mass is always greater than the mass of N$_2$, so constituents with mass less than N$_2$ are always subject to escape when hydrodynamic escape is occurring. Ar, Kr and Xe all have masses greater than N$_2$, but the proposed escape rate of N$_2$ is so high that the critical mass is several times greater than the mass of Xe and the last term in Eq. \ref{eq8} has a value of 1.0. This means that the escape rate of the noble gases relative to the escape rate of N$_2$, under hydrodynamic escape conditions, is equivalent to the abundance of the noble gas relative to N$_2$ as suggested in Eq. \ref{eq8}. Therefore, at the highest possible escape rate for the noble gases, the abundance of the noble gas relative to N$_2$ will remain constant.  In the case of escape rates that are subsonic, escape of the noble gases will be far less efficient due to insufficient energy to drag the heavier molecules from the atmosphere.  As a result, the noble gas escape rate relative to N$_2$'s rate of escape will be less than the abundance of the noble gases relative to N$_2$ in the atmosphere.  Over time, the differential escape rates will lead to an increase in the noble gas abundance in Pluto's atmosphere.  This means that without trapping of Ar by clathrates, escape over an extended period of time (millions of years) will lead to a noble gas abundance that either remains constant or increases with time. Therefore, a depletion of Ar, Kr and Xe in the atmosphere can best be explained by trapping of the noble gases in clathrates.

\section{Discussion}
\label{discus}

Using the noble gas abundances given in Table \ref{atm} and the fitting laws of sublimation laboratory data proposed by Fray \& Schmitt (2009), we find that the equilibrium temperatures of Ar, Kr and Xe pure ices are $\sim$39.5, 42.5 and 55.3 K at the surface pressure of 2.4 Pa, respectively. Given Pluto's average surface temperature ($\sim$50 K), Xe appears stable on the surface in present-day conditions and the condensation of this noble gas should also induce a decrease of its atmospheric abundance. This implies that there is no way to disentangle Xe's previous trapping in clathrate versus its simple condensation. This is fortunately not the case for Ar and Kr, so the measurement of at least one of these two noble gases is critical for testing the scenario of clathrate trapping. The UV spectrometer Alice aboard the New Horizons spacecraft will be sensitive enough to detect an argon abundance that is $\sim$ 10 times smaller than protosolar Ar/N in some circumstances (see appendix \ref{A}).

Our scenario of the noble gas trapping in clathrates on Pluto is motivated by the fact that the same interpretation was provided in the case of Titan in order to account for its observed noble gas deficiency (Osegovic \& Max 2005; Thomas et al. 2007, 2008; Mousis et al. 2011). The situation here is much more favorable than in the case of Titan since the clathrate layer required on Pluto's surface is $\sim$10$^4$ times thinner. This is essentially due to the difference between the surface pressures since the two atmospheres are both dominated by N$_2$. 

{Note that our model considers only the trapping of the noble gases that were present in the early atmosphere. It does not exclude the possibility that the bulk of these noble gases is still in the interior of Pluto. However, if there is outgassing, the noble gases released in the atmosphere should be trapped as well by clathrates. Our calculations are also valid irrespective of the source (primordial or radiogenic) of the noble gases potentially present in Pluto's atmosphere. Indeed, since Pluto is $\sim$half rock, a significant part of the existing argon could result from the radiogenic decay of potassium-40.}

Interestingly, similar calculations have been performed in the case of Triton and they lead to conclusions as favorable as in the case of Pluto. However, the surface temperature of Triton, $\sim$38 K (Tryka et al. 1994), is much lower than Pluto's. It is low enough that the condensation of the three noble gases as pure ices on the surface would largely remove them from the atmosphere irrespective of their sequestration in clathrate. Nonetheless, a mass spectrometer directly sampling Triton's atmosphere could in principle detect the atmospheric abundances of all three noble gases to assess if they were consistent with coexistent surface ices. Since the atmospheric abundances in coexistence with crustal clathrate are orders of magnitude smaller (probably too small for detection), this provides an eventual test of the hypothesis in the case of a future mission to the Neptune system.

\section{Conclusions}
\label{cls}

By considering an atmospheric composition close to that of today's Pluto and a broad range of surface pressures, we find that Ar, Kr and Xe can be efficiently trapped in clathrates if they formed at the surface. The formation of noble gas-rich clathrates on Pluto could then induce a strong decrease of their initial atmospheric abundances. A clathrate thickness {of order of a few centimeters} globally averaged on the planet is indeed enough to trap Ar, Kr and Xe {irrespective of the clathrate structure,} if they were in protosolar proportions in the early Pluto's atmosphere. We suggest that the measurement of the Ar abundance by the Alice ultraviolet spectrometer (Stern et al. 2008) aboard the New {Horizons} spacecraft during Pluto's flyby should {provide} a test of the validity of our scenario.

\section*{Acknowledgements}

O. Mousis acknowledges support from CNES. J.I. Lunine was supported by a contract from JPL under the Distinguished Visiting Scientist program. J. H. Waite acknowledges support from NASA Rosetta funding.

\appendix

\section{Estimating Argon Airglow Emission for the New Horizons Mission}
\label{A}

With the nominal Pluto encounter of the New Horizons mission fully planned, we wish to estimate the amount of Ar I that can be detected by the Alice UVS instrument during approach. There will be 11 Alice observations at close range on approach dedicated to measuring extended airglow emission, with integrations times ranging from 360 to 6900 seconds each, for a total of approximately 6 hours of integration. The midpoint distances from Pluto for these observations range from 1 $\times$ 10$^6$ km to 3 $\times$ 10$^5$ km. We use the planned observation times and their corresponding midpoint distances, as well as the Alice effective area at the Ar I doublet (104.8, 106.7 nm), to estimate the signal to noise ratio (SNR) for a given Ar I mixing ratio. 

The Ar I brightness was determined using the Atmospheric Ultraviolet Radiance Integrated Code (hereafter AURIC) (Strickland et al. 1999). This was run with Ar I abundances of 0.1\%, 0.3\%, 1\%, 3\%, and 10\%, to produce volume production rates from photoelectron impact and photoexcitation processes. ``Model 2'' densities from Krasnopolsky \& Cruikshank (1999) were used for N$_2$ and CH$_4$ in the model atmosphere (7.1 $\times$ 10$^{13}$ cm$^{-3}$ and 6.4 $\times$ 10$^{11}$ cm$^{-3}$, respectively), with laboratory measured cross sections and TIMED/SEE solar spectral irradiance for excitation. The Ar I brightnesses fed into the Alice UVS model thus are the emergent intensities upwelling through the N$_2$/CH$_4$ atmosphere to space.

\begin{figure}
\begin{center}
\includegraphics[angle=90,width=8cm]{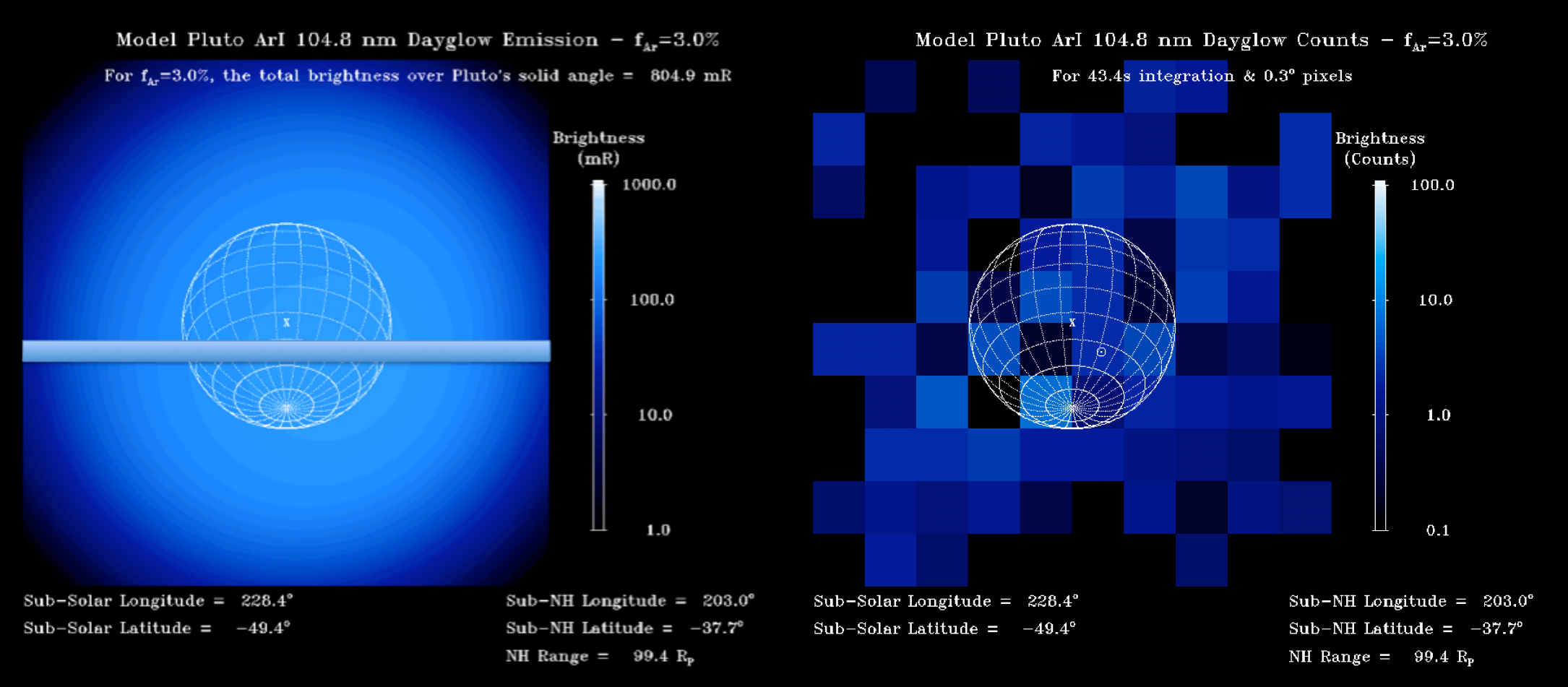}
\caption{Left: Simulated Ar I 104.8 nm airglow emission with a 3\% mixing ratio. Right: Count rate images for Alice observation with a 3\% mixing ratio.} 
\label{airglow}
\end{center}
\end{figure}

\begin{figure}
\begin{center}
\resizebox{\hsize}{!}{\includegraphics[angle=0]{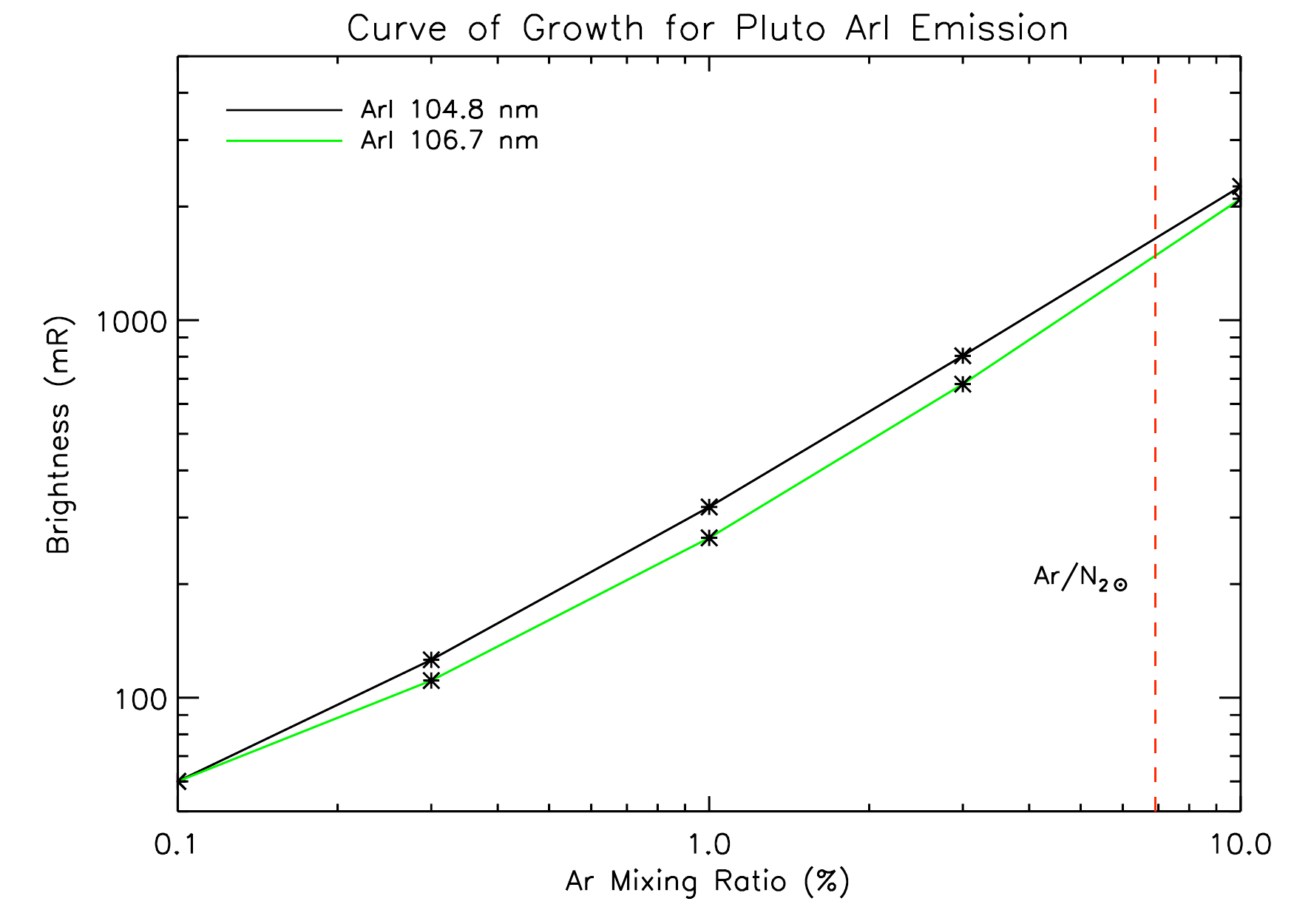}}
\caption{Emergent Ar I 104.8 and 106.7 nm brightness as a function of mixing ratio.} 
\label{ArI}
\end{center}
\end{figure}

Figure \ref{airglow} shows simulated brightness images for one of the approach observations, along with the observation
coordinates. Figure \ref{ArI} shows the curve of growth made with the brightness estimates for each Ar I abundance simulated. We use this brightness relation to calculate the SNR obtained with the Alice airglow observations. The total counts obtained ($C$ in photons) in the Alice slit is just:

\begin{equation}
C = \frac{10^6}{4 \pi}~R~A~\sum t_i \Omega_i
\end{equation}

\noindent where $R$ is the brightness (Rayleighs), and $A$ is the effective area (0.1 cm$^2$ at the Ar I doublet). Since each observation differs in length ($t_i$ in seconds) and is performed at different distances (leading to different solid angles), we sum the photons piecewise. $\Omega_i$ is the effective solid angle of the emitting object visible in the slit. The different exposure times ($t_i$), midpoint distances from Pluto ($d_i$), and slit-filling solid angle for each observation are listed in Table \ref{data}.

\begin{figure}
\begin{center}
\resizebox{\hsize}{!}{\includegraphics[angle=0]{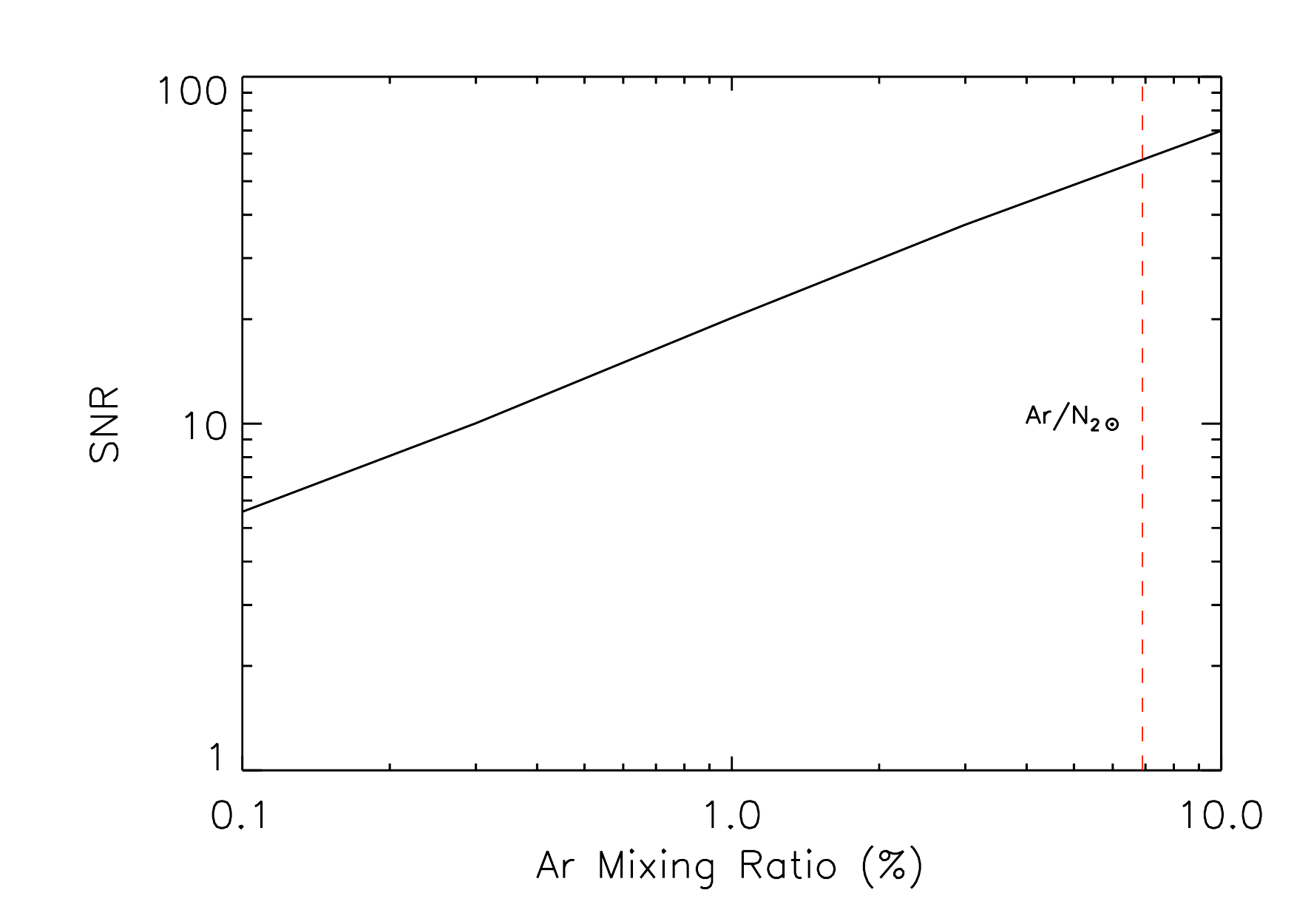}}
\caption{Signal to noise ratio for the added Ar I 104.8 and 106.7 nm line emission vs. mixing ratio.}
\label{mix}
\end{center}
\end{figure}

For our SNR estimates we consider only the emission from the area of Pluto's solid disk. We assume a detector background dark rate of 0.02 counts per spatial/spectral element (as measured in flight) and the AURIC brightness estimates of the 104.8 and 106.7 nm Ar I emission lines to calculate the SNR for a given mixing ratio, shown in Figure \ref{mix}. 

With these calculations we conclude that in 6 hours of observing on final approach to Pluto, the Alice instrument should be able to detect Argon emission in Pluto's atmosphere even at lower than solar abundances. {Assuming a number abundance of 0.25 ppm for Ar I and 36 ppm for N$_2$, i.e., at one tenth solar abundance, the SNR will be $\sim$16. Unfortunately, the measurements of Kr and Xe abundances in the atmosphere of Pluto is beyond the instrument's capabilities.}

\clearpage

\begin{table*}[h]
\centering \caption{Parameters for Kihara and Lennard-Jones potentials}
\begin{tabular}{clccc}
\hline \hline
Molecule   & $\sigma_{K-W}$ (\AA)& $\epsilon~_{K-W}/k_B$ (K)& $a_{K-W}$ (\AA) 	& Reference		\\
\hline
N$_2$			& 3.0993		& 133.13		& 0.3526	&	{Herri \& Chassefi{\`e}re (2012)}		\\
Ar              			& 2.9434     	& 174.14     	& 0.184 	&	{Herri \& Chassefi{\`e}re (2012)}		\\
Kr             			& 2.9739     	& 198.34     	& 0.230 	&	Parrish \& Prausnitz (1972)		\\
Xe              		& 3.32968     	& 193.708     	& 0.2357 	&	Sloan \& Koh (2008)				\\
\hline
\end{tabular}\\
$\sigma_{K-W}$ is the Lennard-Jones diameter, $\epsilon_{K-W}$ is the depth of the potential well, and $a_{K-W}$ is the radius of the impenetrable core, for the guest-water pairs.
\label{Kihara}
\end{table*}

\clearpage

\begin{table}[h]
\centering \caption{Parameters of the dissociation curves for various single guest clathrate hydrates. A is dimensionless and B is in K. {Constants for Kr and Xe are given for an Arrhenius law making use of Napierian logarithm.}}
\begin{tabular}{lcc}
\hline \hline
Molecule        		& $A$           		& $B$  		\\
\hline
N$_2$           		& 9.86          		& $-$728.58  	\\
Ar              			& 9.34          		& $-$648.79  	\\
Kr              			& 22.3934          	& $-$2237.82  	\\
Xe              		& 16.62          		& $-$3159	 \\
\hline
\end{tabular}
\label{tab:coeff_pression}
\end{table}

\clearpage

\begin{table}[h]
\centering \caption{Assumed composition of Pluto's atmosphere at the ground level. Noble gas abundances relative to N$_2$ are assumed to be protosolar (Asplund et al. 2009).}
\begin{tabular}{lcc}
\hline \hline
Species $K$			& Mole fraction $x_K$  			\\
\hline
N$_2$				& 9.31 $\times$ 10$^{-1}$			\\
Ar              				& 6.92 $\times$ 10$^{-2}$   		 \\
Kr             				& 4.90 $\times$ 10$^{-5}$   		\\
Xe              			& 4.79 $\times$ 10$^{-6}$   		\\
\hline
\end{tabular}
\label{atm}
\end{table}

\clearpage

\begin{table}[h]
\centering \caption{Relevant parameters of each airglow observation.}
\begin{tabular}{lccc}
\hline 
\hline
Observation			& $d_i$ (km)	& $t_i$ (s)		& $\Omega_i$ ($\mu$rad)	\\ 
\hline
PC--Airglow--Fill--0 		& 1116455 	& 6900 		& 3.21					\\ 
PC--Airglow--Fill--2 		& 819350 		& 600 		& 4.70					\\ 
PC--Airglow--Fill--2 		& 809012 		& 600 		& 4.77					\\ 
PC--Airglow--Fill--2 		& 798261 		& 600 		& 4.84					\\ 
PC--Airglow--Fill--2 		& 788336 		& 600 		& 4.91					\\ 
PC--Airglow--Fill--2 		& 780204 		& 360 		& 4.97					\\ 
PC--Airglow--Appr--1a 	& 611833 		& 1200 		& 6.49					\\ 
PC--Airglow--Appr--1b 	& 575456 		& 600 		& 6.94					\\ 
PC--Airglow--Appr--2 	& 525813 		& 4200 		& 7.63					\\ 
PC--Airglow--Appr--3 	& 408242 		& 3600 		& 9.93					\\ 
PC--Airglow--Appr--4 	& 318118 		& 3000 		& 12.83					\\ 
\hline
\end{tabular}
\label{data}
\end{table} 

\clearpage

\clearpage

\clearpage

\clearpage

\clearpage

\end{document}